# Junk News Bubbles
## Modelling the Rise and Fall of Attention in Online Arenas

Maria Castaldo, Tommaso Venturini, Paolo Frasca, Floriana Gargiulo

*In this paper, we present a type of media disorder which we call "junk news bubbles" and which derives from the effort invested by online platforms and their users to identify and circulate contents with rising popularity. Such emphasis on trending matters, we claim, can have two detrimental effects on public debates: first, it shortens the amount of time available to discuss each matter; second it increases the ephemeral concentration of media attention. We provide a formal description of the dynamic of junk news bubbles, through a mathematical exploration of the famous "public arenas model" developed by Hilgartner and Bosk in 1988. Our objective is to describe the dynamics of the junk news bubbles as precisely as possible to facilitate its further investigation with empirical data.*[1]

## Conceptualizing junk news bubbles

Much has been written in the last years about online media and the threat of "selective exposure" (Sears & Freedman, 1967), "echo chambers" (Garrett, 2009) and "filter bubbles" (Pariser, 2011). In various ways, all these notions point to the risk that the growing availability of information and the perfecting of filtering and recommendation technologies may create a "situation in which thousands or perhaps millions or even tens of millions of people are mainly listening to louder echoes of their own voices" (Sunstein, 2001, p.16). Reviving a long tradition of homophily and segregation models (Schelling, 1971), this idea has sparked much interest in computational sociology (cf., among others, Colleoni et al., 2014; Barberá et al., 2015; Quattrociocchi et al., 2016; Geschke et al., 2019) despite evidence that its impacts may be overestimated (cf. Bakshy et al., 2015; Flaxman et al., 2016; Boxell, et al., 2017; Dubois, 2018).

Less attention and computational efforts have been dedicated to a different type of media bubbles, whose danger comes not from the fragmentation, but from the ephemeral concentration of public attention. In a previous paper (Venturini, 2019), we proposed the term "junk news" to expand the notion of "fake news" beyond its excessive focus on deceitfulness (cf., among others, Zuckerman, 2017; Benkler et al., 2018; Gray et al. 2020). Besides false contents disguised as mainstream news and explicitly directed at deceiving their receivers, media scholars should worry about the avalanche of memes, click-baits, trolling provocations and other forms of ephemeral distractions that prevent online audiences from engaging in a thoughtful public debate. This type of "information disorders" (Wardle & Derakhshan, 2017) cannot be defined on the basis of its content or its style (as in manual and automated fact-checking, cf. Graves, 2016 and Ciampaglia et al., 2015) nor on the basis of the way in which it polarizes the debate (as in echo chambers and filter bubbles). Instead, ephemeral distractions are distinguished by a feature often neglected when considering online misinformation – their temporal profile. Like junk food, junk contents tantalize without ever satiating. Like echo chambers, they can be described as bubbles, although not the sense of a secluded informational space, but rather as market bubbles: speculative waves that destroy rather than create richness in public debate.

If the dangers of junk news bubbles are neglected by current media research, it is not because they are lesser than that of fake news and filter bubbles, but because this particular information disorder still lacks the precise conceptualization that supports the research on fact-checking and polarization. To outline such a conceptualization, we propose to define junk news bubbles as an *adverse media dynamic in which a large share of public attention is captured by items that are incapable of sustaining it for a long time*. Both elements of this definition are crucial. Popular stories and even viral contents are not necessarily junk news, no matter how quickly and largely they spread in online networks (Jenkins et al. 2013). To qualify as junk news bubbles, contents must fade away as quickly as they rose, so that they distract public debate rather than nourishing it. Note that our definition is agnostic about the quality of junk contents. Even a patent piece of misinformation such as the infamous claim that "Brexit will make available 350 million pounds per week for the NHS" plastered onto a red bus during the UK-EU referendum campaign, can end up generating productive discussions if it sticks long enough in the public debate (Marres, 2008). Vice versa, newsworthy stories cannot contribute to democratic conversation if they are too quickly pushed out of the public agenda. In other words, the notion of junk news bubbles applies less to specific pieces of content, than to a general acceleration of online attention cycles.

[1] This work has been partially supported by MIAI@Grenoble Alpes (ANR-19-P3IA-0003) and by CNRS via project 80|PRIME "Disorders of Online Media (DOOM)".

Central in the '70s and '80s, the question of "attention cycles" (Downs, 1972) has lost steam in current media research because of the advent of digital technologies and the extension of the media system that they brought with them. Because of this extension, the question of the occupation of public debate has begun to be formulated in spatial rather than in temporal terms (i.e. where something is discussed rather than when). Temporal dynamics, however, remains crucial for, as in the words of McLuhan, "the 'message' of any medium or technology is the change of scale or pace or pattern that it introduces into human affairs... amplif[ying] or accelerate[ing] existing processes" (McLuhan, 1964, p.8). As noted by scholars working on the attention economy (Lanham, 20066; Terranova, 2012; Crogan & Kinsley, 2012), digital technologies are particularly inclined to amplify "media hypes" (Vasterman, 2005) and to concentrate public attention on widespread but ephemeral trends.

Our argument echoes McCombs's plea (McCombs, 2005) not to desert the research on "agenda setting" (McCombs & Shaw, 1972; McCombs, 2004) on the grounds of online attention being scattered in a myriad of different channels each with its own agenda. Precisely because of this overabundance of sources, digital media have since their inception channeled their flows through increasingly sophisticated "Engines of Order" (Rieder, 2020) constantly ranking – and updating the rank – of their contents (Cardon, 2005; Cardon *et al.*, 2018). This is particularly true of social media platforms, whose size could not be sustained without infrastructures to sieve through millions of contents produced every day and whose business model as marketplaces of ideas entails a relentless circulation of new trending topics (Webster, 2014). This unremitting production of trends is an inherent feature of recommendation algorithms, as candidly admitted by YouTube engineers: "in addition to the first-order effect of simply recommending new videos that users want to watch, there is a critical secondary phenomenon of boot-strapping and propagating viral content" (Covington *et al.*, 2016 p.193, see also Zhao *et al.*, 2019). Algorithms, however, are not the only component of online attention infrastructures and not the only actors at play. A similar push for trendiness comes from the interaction between the practices of social media users, which are increasingly driven by micro-celebrity strategies (Marwick & Boyd, 2011; Khamis et al. 2017), and the graphic interface of platforms and apps, which are partial to ephemeral visibility and vanity metrics (Rogers, 2018).

The term "junk news bubbles", thus, designates less a specific type of content (as in the case of misinformation or "fake news") than a general tendency of media systems to privilege trendy items and thus create an extremely turbulent and superficial public debate. False and low quality stories are the type of content that thrives most directly on the incentives of trendiness, but to some extent all digital conversations are influenced by this acceleration. In the paper where we first proposed the term "junk news" (Venturini, 2019), we discuss in more detail the political economy of online acceleration and the way in which it influences the circulation of news. In this paper, our goal is to provide a *formal* description of these attention dynamics in order to encourage their further empirical study. With a few remarkable exceptions (see in particular Leskovec, Backstrom & Lars, 2009 and Lorenz-Spreen et al., 2019), no large-scale research has been devoted to attention cycles, despite the growing availability of traces produced by digital media (Lazer et al., 2009; Latour et al, 2012; Venturini, Jensen & Latour, 2015).

To facilitate such research, we propose a mathematical formalization of one of the most influential accounts of attention dynamics: the "public arenas model" introduced in 1988 by Stephen Hilgartner and Charles Bosk. Despite its clarity and insightfulness, H&B's framework has so far found no mathematical formalization for its complexity and lack of formal description. In this paper, we streamline H&B's model focusing on the rise and fall of attention matters (and ignoring the linkages across different arenas and the actors within each arena). Doing so we propose a ready-to-test (*prêt-à-tester*) version of H&B s model hoping that it will encourage further empirical investigation on junk news bubbles. Ours is a case of "toy model" (Reutlinger *et al.*, 2018), whose function is not to be applied or fitted to empirical data nor to offer an accurate description of the phenomenon that it presents, but to help in defining it and setting the conceptual bases for its future study.

## Model description

**(a)** The first ingredient of our model is a **population of "matters of attention"** (or "social problems" as in H&B original formulation) defined self-referentially as the *entities that compete to capture public attention*. The non-essential nature of this definition is crucial for H&B, who contend that "social problems are projections of collective sentiments rather than simple mirrors of objective conditions" (H&B p.54). In other words, matters of attention are defined by their visibility and not the other way around ("we define a social problem as a putative condition or situation that is labeled a problem in the arenas of public discourse and action" p.55). Three corollaries descend from this non-essentialist definition:

- First, *all attention matters are equal before our model* and their rise and fall depend exclusively on the competition between them and not on any substantial features ("social problems exist in relation to other social problems" p.55).
- Second, *our model focuses on attention dynamics internal to media arenas*, deliberately disregarding the influence of exogenous shocks. This does not mean, of course, that these shocks do not exist (clearly the breaking of a war or of an earthquake will command attention in all attention arenas). Yet, their influence is both obvious and insufficient to account for all media dynamics ("if a situation becomes defined as a social problem, it does not necessarily mean that objective conditions have worsened. Similarly, if a problem disappears from public discourse, it does not necessarily imply that the situation has improved" p.58). This is particularly true of the kind of junk news we are interested in, which may occasionally surf the drama of external events, but is more often entirely self-referential. For these reasons, exogenous shocks are deliberately excluded from our model (but empirical applications should, of course, control for them).
- Third and similarly to H&B framework, *our model can be applied to different media and at different scales*. Attentions matters are broadly defined as recognizable units of content in a particular forum of collective debate (the attention arena). Examples could be different videos in a given YouTube channel or different threads in a given Reddit subreddit. To be sure, we are not promising that our model will fit all media debate but inviting scholars to test it empirically on different phenomena to determine to which it can be fruitfully applied.

**(b)** The second ingredient of our model are two **competition mechanisms** that favor some attention matters over others. The four different "principles of selection" distinguished by H&B find in our model a formalization in two main mechanisms:

- *Exogenous influences*. Three of the four "principles of selection" distinguished by H&B, "drama" (pp.61-62), "culture and politics" (H&B p.64) and "organizational characteristics" (pp. 65,66) are rendered in a deliberately coarse way in our model. The dramatic value of attention matters as well as the way in which they resonate with the general culture or with the specific organization of the medium are important, but their influence falls outside the self-induced media dynamics that constitute the focus of our model. In our formalization, the influence of these features is thus rendered as a noise which randomly increases or decreases the visibility of each item at each iteration. This solution allows for accounting for this type of influence (and to explore the effect of its variation) under the assumption that its specific nature does not affect the dynamic of junk news bubbles.

- *Endogenous trending*. The last selection principle identified by H&B, "novelty and saturation", is crucial to our model. At each iteration, the model increases or decreases the visibility of each matter, repeating its previous variation, *multiplied by a parameter that accelerates or decelerates such variation*. The model therefore rewards rising items and penalizes declining ones. This mechanism works as a *Matthew effect* (Merton, 1968 and Newman, 2001), but a *dynamic one* which rewards not the most visible matters, but the ones that have increased the most since the previous iteration. This boosting of trendiness is consistent with the way in which online platforms "emphasiz[e] novelty and timeliness… [by] identifying unprecedented surges of activity" and "reward[ing] popularity with visibility" (Gillespie, 2016, p.55&60). Such partiality for trendiness is characteristic of both social media and their users, in a sociotechnical loop in which the visibility granted by platform algorithms both *depends on* and *is influenced by* the number of views generated by different contents.

**(c)** The third ingredient of our model are the **attention boundaries**. At each iteration, *after* adding (or subtracting) to each attention matter its random variation and its trending acceleration, the model *corrects* the potential visibility of each item to make sure that it remains within two inflexible boundaries:

- *Lower boundary: exclusion of negative visibility*. Because it is impossible to conceptualize such a thing as negative attention, when noise or acceleration push the visibility of a matter of attention below zero, the item is removed from the arena and replaced with a new one with null initial visibility. Because a new attention matter can enter the arena only when an old one leaves it, the number of items in the model remains fixed (but some items can have visibility equal to zero).
- *Upper boundary: saturation of the attention capacity*. After having applied noise and acceleration and corrected for negative attention, the model divides the potential visibility of each item by the sum of the potential visibilities of all items. This normalization makes sure that the sum of all computed visibilities remains equal to one. This boundary implements a key ingredient of H&B framework, the idea that each debate arena has a **fixed attention capacity** (or "carrying capacity", in H&B terms). The fixity of the global "carrying capacity" is crucial to ensure that our model does not converge to a trivial winner-takes-all equilibrium. While raising attention matters are pushed to an increasing visibility by their trendiness, they all end up reaching a point where they exhausted their potential for growth, begin to slow down and are penalized by competition mechanisms.

  The inelasticity of attention capacity also ensures that the visibility gained by one matter of attention is always lost by some other so that "the ascendance of one social problem will… be accompanied by the decline of one or more others" (H&B p.61). While we are, of course, aware that public attention fluctuates with circadian and professional rhythms, we believe that these cyclical fluctuations can be discounted for the sake of simplicity. Following H&B, we think that good reasons for a fixed attention capacity can be found in the limited staging capacity of media ("the prime space and prime time for presenting problems publicly are quite limited" p.59) and, more importantly, in the limited capacity of the public to attend to public ("members of the public are limited not only by the amount of time and money they can devote to social issues, but also by the amount of 'surplus compassion' they can muster for causes beyond the usual immediate concerns" p.59). This second element is crucial to understand why the assumption of a limited carrying capacity remains relevant for online media even if digital technologies removed most of the barriers of conventional news gatekeeping (Shoemaker & Vos, 2009). As noted by Ray Maratea in relation to the visibility in the blogosphere:

> "Although the Internet may provide an infinite carrying capacity and make the claims-making process more efficient, it cannot resolve the problem that audiences have limited amounts of time and attention to focus on various social problems. Furthermore, the blogosphere has developed according to a hierarchical structure, meaning readers and traditional journalists largely concentrate their attention on a relatively small number of well-known blogs. In short, the development of the blogosphere has the potential to dynamically change the claims-making process, but Hilgartner and Bosk's public arenas model remains vital to understanding the rise and fall of social problems in the new media age" (Maratea, 2010, p. 140).

In fact, by eroding the boundaries between news and entertainment (Prior, 2005), digital convergence may have created an even harsher "all-out war for the time of an audience that has more choices than at any point in history" (Klein, 2020, p. 279). While others (see for example, Cinelli et al., 2019) takes this competition as a reason for selective exposure and filter bubbles, we believe that they are a crucial ingredient of ephemeral concentration and fake news bubbles. According to the limited capacity model of mediated message processing proposed by Annie Lang (2000), media users can deal with an overabundance of media stimuli by switching to an entertainment – rather than understanding – mode, saving their limited cognitive resources by interacting more superficially with the medium ("running on automatic", Lang, 2000, p. 53).

## Model formulation and parameters

**(a)** We call $x_i$ each item of our **population of matters of attention**, with $i = 1, \ldots, n$, where $N$is the maximum number of items in the population. We call "visibility" or ${\pi^t}_i$ the share of attention captured by $x_i$ at time $t$. By a mechanism explained below, at each timestep, the sum of ${\pi^t}_i$ for all $i$ is fixed and equal to one. This allows to interpret each ${\pi^t}_i$ as the percentage of the total attention captured by each item $i$ at time $t$

**(b)** We model the two **competition mechanisms** as follows:

- *Endogenous trending*. At every timestep $t+1$, the visibility ${\pi^{t+1}}_i$ of each item $i$ is modified by adding to its current visibility ${\pi^t}_i$ a term which repeats its previous variation (i.e. $\Delta\pi = {\pi^t}_i$- ${\pi^{t-1}}_i$) multiplied by a positive factor $\alpha$, which could be interpreted as a boost of trendiness.
- *Exogenous influences*. In our formalization, we render all external influences on media dynamics as a noise ${\varepsilon^t}_i$which increases or decreases the visibility of item $i$ randomly at timestep $t$. The noise ${\varepsilon^t}_i$ is a realization of a normal distribution

  $N(0, \frac{1}{c*n^2})$ with mean = 0 and standard deviation = $\frac{1}{\sqrt{c}n}$

  where *c* is a positive parameter. We can therefore write the potential visibility of each item after the iteration ${p^{t+1}}_i$ as the output of the two above mechanisms as follows:

$$ {p^{t+1}}_i := {\pi^t}_i + \alpha({\pi^t}_i - {\pi^{t-1}}_i) + {\varepsilon^t}_i \qquad (1) $$

**(d)** At each iteration $t$, the potential visibility ${p^{t+1}}_i$ is replaced with its corrected version ${\tilde{p}^{\,t+1}}_i$ to abide by the model's **attention boundaries**:

- *Exclusion of negative visibility*. ${\tilde{p}^{\,t+1}}_i$ equals ${p^{t+1}}_i$ if ${p^{t+1}}_i$is positive. Otherwise it is set to zero. Hence,

$$\tilde{p}^{t+1}{}_i = max(0, p^{t+1}{}_i) \tag{2}$$

- *Saturation of the attention capacity*. The limited capacity of an arena is represented by the constraint of having a fixed sum of popularities at each timestep. Therefore, each visibility is obtained from the non-negative $\tilde{p}^{t+1}{}_i$ by normalization.

$$\pi^{t+1}{}_i = \frac{\tilde{p}^{t+1}{}_i}{\sum_j \tilde{p}^{t+1}{}_j} \tag{3}$$

Initialization. At the first step of the model, the visibility of every $i$ (i.e. $\pi^1{}_i$) is initialized with a random numbers drawn from a uniform distribution between 0 and 1 and normalized to satisfy the constraint $\sum_i \pi^1{}_i = 1$. At the second step, the visibility every $i$ (i.e. $\pi^2{}_i$) is obtained by adding to $\pi^1{}_i$ a noise $\varepsilon^t{}_i$ drawn from the normal distribution $N(0, \frac{1}{c*n^2})$ and normalizing. After the first two steps, the dynamics is self-sustained by evaluating equations (1), (2) and (3) at each iteration.

Inspecting the equations above, it is easy to observe that our model has only three parameters:

- $\alpha$, trendiness boost, which decides whether the visibility variation at the previous iteration is amplified at the next one and by how much, is the key parameter of our model. Conceptually, $\alpha$ can be interpreted as the keenness of media algorithms and media users to identify and promote trendy matters of attention. The bigger is $\alpha$, the more important is the role played by trendiness in the sociotechnical choices that influence the visibility of media items. High values of trendiness boost thus simulate the attention dynamics occurring in debate arenas prone to junk news bubbles.
- The other two parameters are
    - $n$, which represent the maximum number of attention matters simultaneously present in the simulation,
    - and $c$, which represent the size of noise, that is to say the importance of exogenous influences.

  Both $n$ and $c$ are used in the realization of noise and, because they appear in the denominator of the distribution that generates noise, the higher they are, the smaller are the variations due to noise.

## Model results and discussion

Despite its simplicity, our model is able to generate patterns comparable with the empirical observations of media systems (Leskovec, Backstrom & Lars, 2009; Lorenz-Spreen, 2019). In particular, our formalization supports the H&B intuition that the "shifting waves of social problems" (H&B p.67) typical of media attention cycle can be explained by the interaction between the push of trendiness and the saturation of the carrying capacity: "if we explore these complex linkages, we find a huge number of positive feedback loops, 'engines', that drive the growth of particular problems. Growth is constrained, however, by the negative feedback produced by the finite carrying capacities of the public arenas, by competition among problems for attention, and by the need for continuous novel drama to sustain growth" (H&B p.67).

Previous studies (Weng *et al.*, 2012; Gonçalves *et al.*, 2011; Cattuto *et al., 2007*) considered the role of users' limited attention in media competition assigning users a maximal number of possible interactions (a sort of Dunbar number for individual attention). In most of these models, the fall of popularity is obtained forcing an aging process of media items through an explicit time decay term. This aging process, however, is difficult to justify theoretically and empirically. One of the most original aspects of our model is that it dispenses with this aging process: items' popularity decays naturally through the interplay between the trendiness and the saturation of the overall attention capacity.

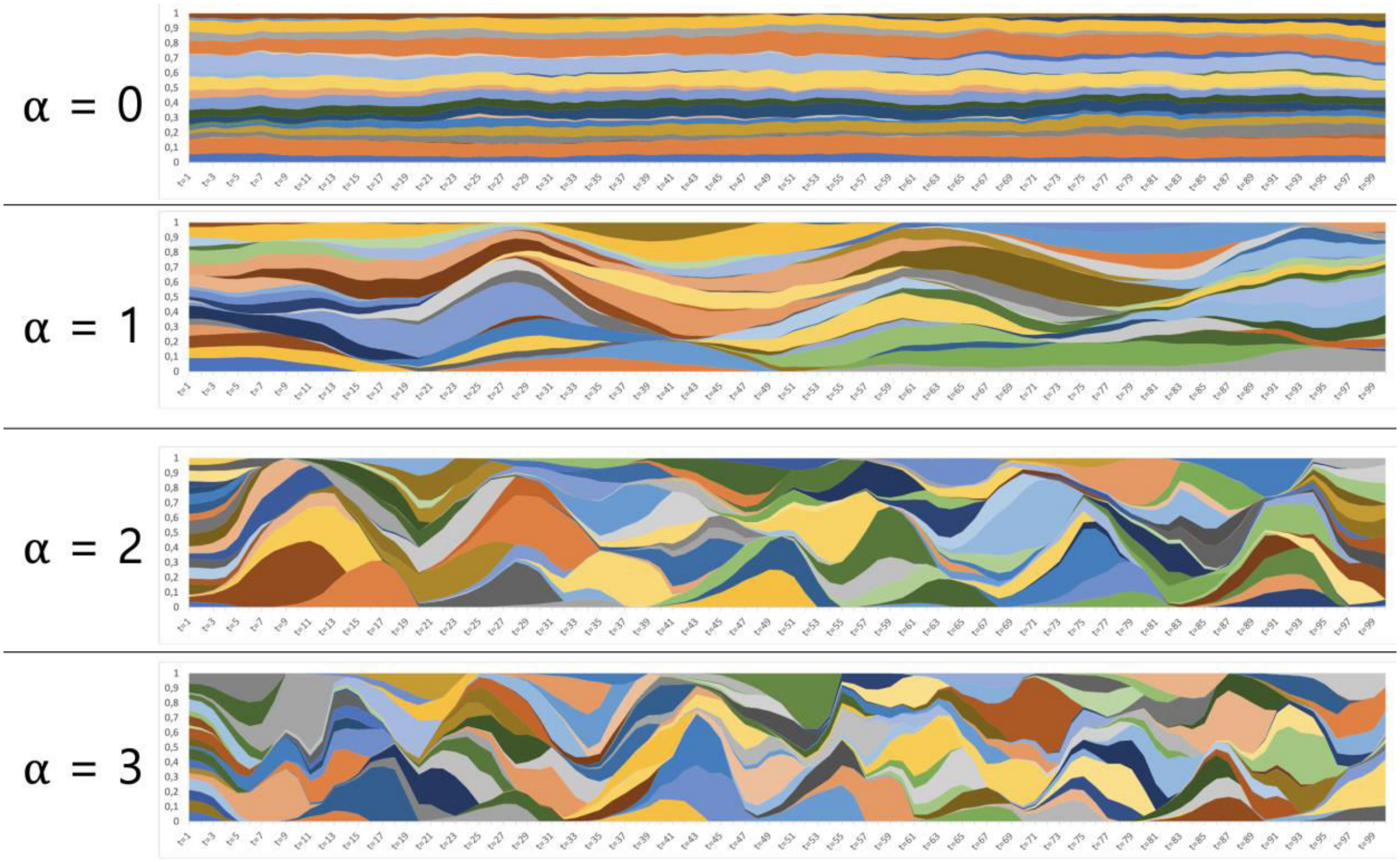

*Figure 1. Evolution of our model for trendiness boost = 0, 1, 2, and 3 (with $n = 20$ and $c = 12$). Each color area corresponds to the attention received by an item. Only the first 100 iterations are shown as the shape of the curves does not change in further iterations*

The comparison between the graphs in fig. 1 suggests that, as the boost of trendiness grows, the rise and fall of attention matters becomes steeper. This relation can be tested by computing the mean steepness of attention curves (the absolute increase or decrease by unit of time) and observing that it increases monotonically with the increase of alpha before reaching a plateau (probably due to the upper and lower constraints on the state and to the impossibility of compressing the width of curve beyond a certain point).

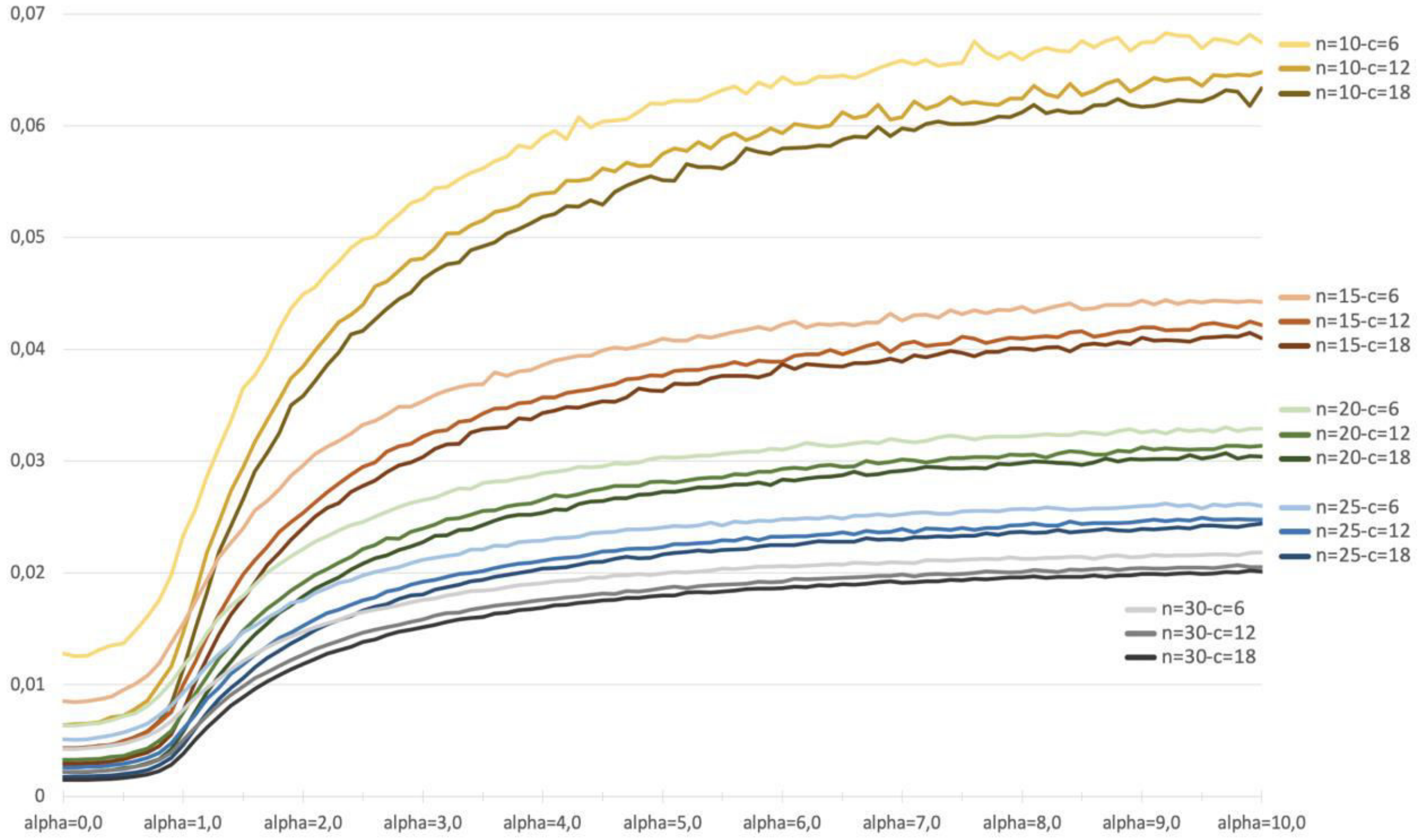

*Figure 2. Mean slope of attention curves as function of the trendiness boost (for different values of n and c)*

Fig.2 confirms that the relation between the steepness of attention curve and trendiness boost is not substantially transformed by the other parameters of our model. The number of attention matters and the importance of exogenous influences shift the position of the curve, but do not change its shape. Also, because both $n$ and $c$ affect the curve in the same way, only $n$ will be explored in the next figures.

Considering together fig. 1 and 2, it is also interesting to notice that trendiness boost increases rise-and-fall steepness by affecting both dimensions of the media cycle: the *height* of attention curves and their *width*. This suggests that junk news bubbles can combine features that may appear contradictory.

- Regardless of the number of items or the level of noise, the stronger is trendiness boost the shorter is the lifecycle of individual attention matters (fig.3a). Remarkably, this is true for all attention matters: even items that reach very high levels of visibility end up falling as quickly as they rose. As a consequence of the shortening of attention weaves, a higher number of matters enter and exit the arena (fig.3b). This may contribute to making platforms more attractive by increasing the dynamism of their offer of information and entertainment.

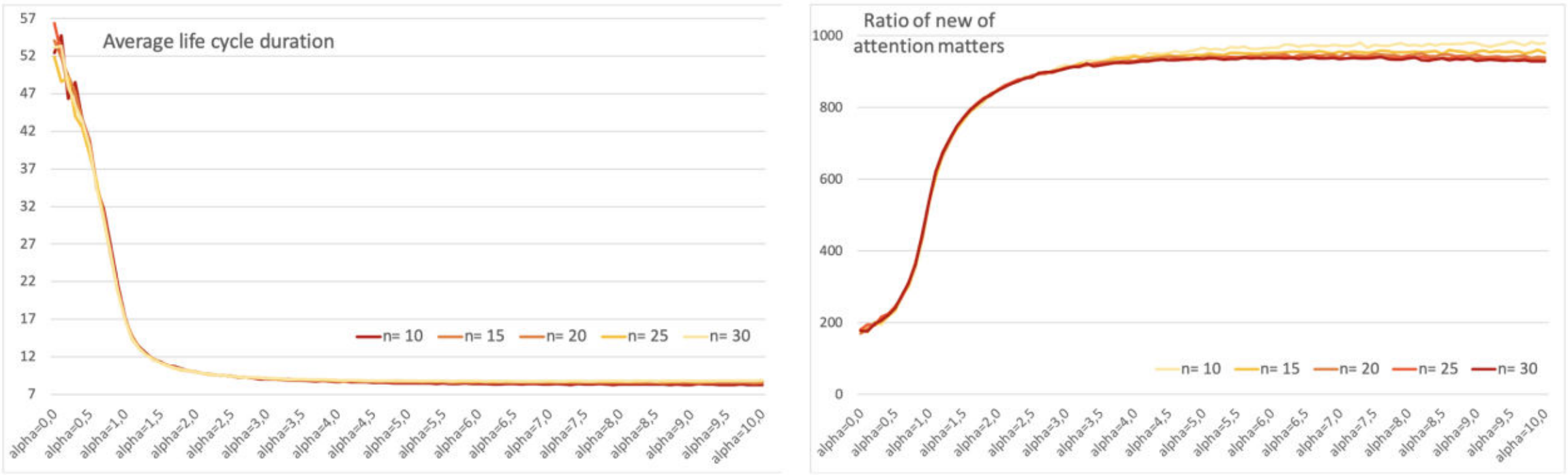

a. b.

*Figure 3. (a) Mean length of attention matters' life cycle and (b) ratio of new attention matters entering the model in its first 10.000 iterations, at the variation of trendiness boost (for different values of n and with c set to 12)*

- On the other hand, higher trendiness boost increases the maximum visibility reached by attention matters (fig. 4a) and, most importantly, amplifies the difference between successful and unsuccessful attention matters, creating a situation in which, at each iteration, most of the available attention is captured by a minority of over-visible items (fig. 4b). "There is a huge 'population' of potential problems-putative situations and conditions that could be conceived of as problems. This population, however, is highly stratified. An extremely small fraction grows into social problems with 'celebrity' status… [while] the vast majority of these putative conditions remain outside or on the extreme edge of public consciousness" (H&B p. 57).

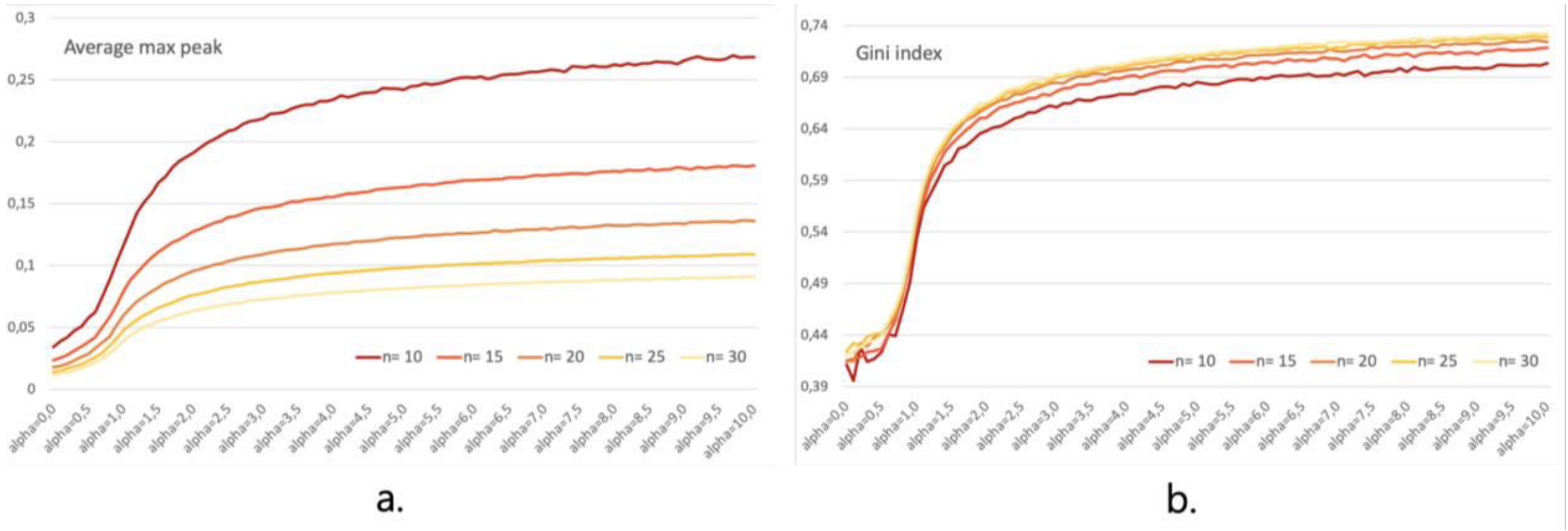

*Figure 4. (a) Mean height of the attention curve peaks and (b) Gini index of attention concentration at each iteration of the model, at the variation of trendiness boost (for different values of n and with c set to 12)*

## Some empirical examples

Being a *toy model*, our mathematical formalization is not meant to serve as a simulation of a real-world dynamic or to be fit to empirical data. Still, simplistic as it is, our toy model does capture some features of the current media system as confirmed, for example, by the results of an empirical investigation of *Accelerating Dynamics of Collective Attention* by Lorenz-Spreen *et al.* (2019). Comparing different media and over several years, the authors found that "the attention associated with individual topics rises and falls with increasing gradients" and "the shifts of collective attention between topics occur more frequently" (p. 1763) suggesting a general increase of trendiness's importance in the media system. They also find that this tendency is stronger for media such as Twitter, Google and Reddit and less pronounced for Wikipedia and the scientific literature, suggesting that the arenas that are more sensitive to trendiness are also the ones that are most exposed to junk news bubbles.

To find other examples of our model's insights, we explored the dynamics of visibility of the French vlogosphere. While the most obvious (and probably the best) operationalization of H&B model would be to consider the whole French YouTube as an arena for the competition of different thematic issues, this would require a huge data collection and automatic identification of themes which exceed the scope of this paper. We thus decided to exploit the *fractal* nature of media systems and to consider each YouTube channel as an arena and each video as an attention matter. As the competition between video is likely to take place across channels rather than within them, this operationalization is far from ideal but, as we will see, sufficient to exemplify our argument.

Starting on December 9, 2019, we recorded the number of views collected every hour by each video published by about thousand YouTube channels active in the French public debate (selection was made by iterating expert review and snowball sampling). For this paper, we limit our analysis to the data collected until March 14, 2020 (in order to exclude the Covid19 lockdown period where we observed slightly different attention dynamics, Castaldo *et al.*, 2020). Since we are exploring in-channel competition, we also restrict our analysis to channels having multiple active videos at the same time, thus focussing on the 60 most active channels on our corpus. In this subcorpus the most active channel is "Europe 1" with 4.291 videos published in the three months of collection and the least active is "Charente Libre" with 216 videos. The most popular channel is "France 24" with 731.950 total views and the least popular "L'Opinion" with 1552 total views. For each of these 60 channels, we calculated the same two metrics computed for in our simulation above: the average life cycle of the videos (defined as the average number of hours necessary for a video in the channel to reach the 95% of the total views collected in the first week) and the average Gini index of hourly concentration.

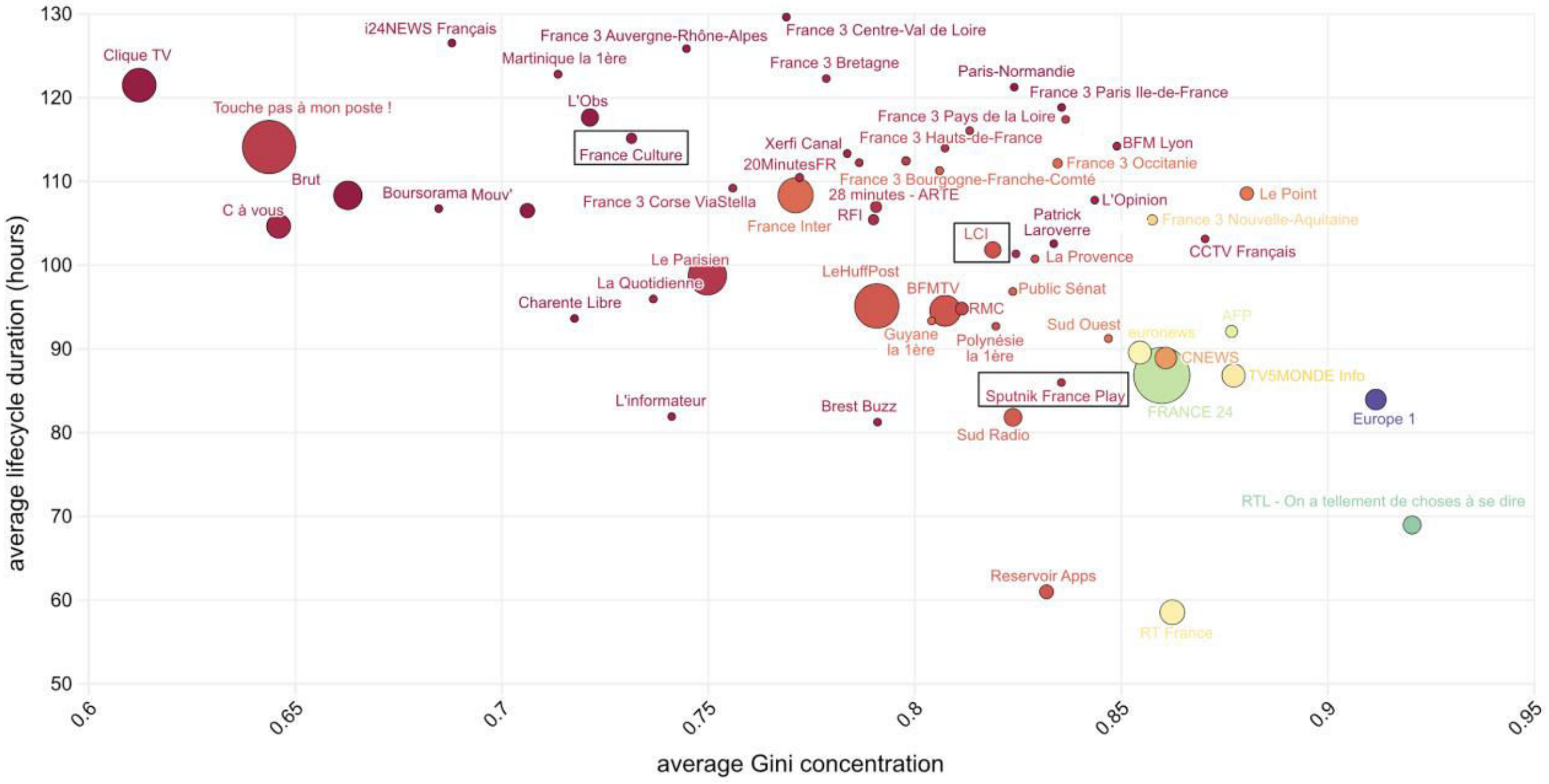

*Figure 5. Scatterplot of the average concentration and average life cycle of the videos of 60 YouTube channels highly active in the French political vlogosphere. The size of the dots corresponds to the total number of views collected by the channel and the color corresponds to the number of published videos (going from dark red for fewer videos, to orange, yellow, green and violet for more videos).*

The scatter plot in fig. 5 confirms the main insight of our model: the fact that short life cycles and high concentration go hand in hand in the distribution of media attention. The distribution on the diagonal of the scatter plot suggests a sort of continuum that goes from channels that publish few videos with a relatively longer life expectancy and a less skewed distribution of attention (top-left), to channels that publishes videos at a much more rapid pace knowing that most will spike and die very quickly, but hoping that some of them will stick (bottom-right). It is tempting to interpret such continuum as a quality VS click-bait, even though many exceptions are manifest in the diagram (e.g. France24, Europe1 and TV5Monde, are news channels of established quality and yet all find themselves in the bottom right corner because of their practices of publishing multiple videos a day).

The scatterplot also allows us to select three chains (boxed in fig. 5) with a similar number of videos and total views, but with very different life spans and concentrations, which we can use as examples of three different attention regimes:

1. "France Culture" (276 videos, 34.694 views) with long life cycles and low concentration, is a renowned public radio, appreciated for the quality of its contents and the seriousness of its debate.
2. "LCI" or "La Chaîne Info" (738 videos, 54.198 views) with medium life cycles and medium concentration, is an established live-news broadcasting service by French public television channel.
3. "Sputnik France" (466 videos, 32.694 views) with short life cycles and high concentration, is a Russian state-owned news agency, often criticized for spreading misinformation and low quality stories.

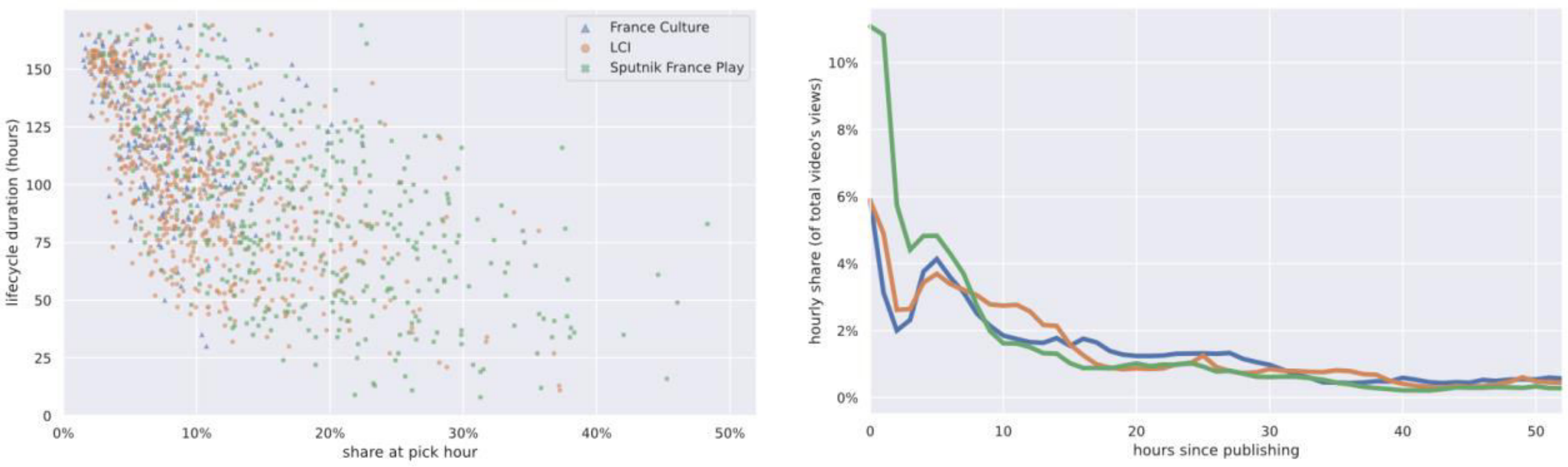

*Figure 6. Left: Scatterplot of the concentration (share at peak hours) and average life cycle of all the videos published by three selected channels (France Culture, LCI and Sputnik France Play). Right: Average temporal profile of the videos of the three channels.*

To highlight the difference between the three attention regimes, fig. 6 (left) plots all of their videos for the duration of their lifecycle (the number of hours needed to reach the 95% of the total views collected in the first week) and the share of visibility over the channel that they obtained at their peak hour (the hour in which they collected most of their views). At their peak hour, videos published by France Culture rarely captured over 10% of the total attention for the channel, but many of them remain alive through most of their first week. The videos by Sputnik France often reach shares of 20% or more, but rarely survive after the first two or three days from their publication. The videos from LCI are somewhere in between. The difference is clearer when observing the average trajectory of visibility for the videos of the three channels for the first two days after their publication (fig. 6 right). While the average Sputnik's video makes most of its views in its first hours and then declines extremely quickly, the average France Culture's video has a slower start, but a thicker visibility tail (LCI presents once again an intermediate profile).

Beside focussing on individual channels, it would be interesting to investigate if, over a longer period of time, it is possible to observe a general shifting down and to the right of the media system as a whole. Unfortunately the data at our disposal do not yet allow such an inquiry.

## Conclusion

Taken alone, none of the consequences of junk news bubbles highlighted by our model is particularly surprising: being an acceleration, trendiness predictably shortens the lifespan of attention matters and, being a positive feedback, it increases their maximum visibility. Their combination, however, is remarkable as it creates a *shoaling* of attention waves which reduces the width *and* increases the height of attention curves. Debate arenas characterized by a stronger focus on trendiness may therefore end up displaying a syncopated rhythm of attention that is at the same time *increasingly dispersed and increasingly concentrated* (as one can easily observe, for example, in YouTube channels or subreddits devoted to buzzing news, memes and viral contents). Junk news bubbles are characterized by the same attention skewness of Boydstun's "media storms" (Boydstun *et al.*, 2014), but not by the same persistence in time. As such, they are particularly worrying because they take attention away from other discussions (because of their skewness), without producing the heightened public awareness created by media storm (because of their ephemerality).

While evidence exists that the syncopation described by our model can be found in numerous online platforms (Wu & Huberman, 2007; Crane & Sornette, 2008; Yang & Leskovec, 2011; Castillo et al., 2014; Bandari et al., 2015), little empirical research has been carried out on the consequences of such attention regime. The risks of distraction related to screens and online media have been decried at the individual and cultural level (cf. for instance, Goldhaber, 1997; Hassan, 2011; Citton, 2014; Crawford, 2015), but hardly investigated through the records increasingly made available by digital platforms themselves. This paper hopes to facilitate such line research by providing a formal description of a distracted attention regime: a situation in which attention waves becomes both higher and narrower and in which public debate is trapped in a continual succession of hot button issues.

Such a situation, arguably, is not particularly propitious to quality. While our model defines junk news bubbles independently from their content value, we suspect this attention regime to be associated with misinformation and poor quality. A regime in which visibility is granted and withdrawn with great rapidity unsurprisingly favours *click-baity* content designed to catch the attention more than to retain it. This observation may explain why, in political discourse, traditional propaganda is increasingly replaced by political trolling aimed at drowning opponents' discourse in noise (Jack, 2017; Flores-Saviaga et al., 2018) or simply to monetize political outrage (Braun & Eklund, 2019).

Being a simplified formalization of a relatively abstract framework, our mathematical model does not allow substantial claims about actual attention dynamics. It allows, however, to advance a precise hypothesis about the junk news bubbles and their detrimental effects on public debate: the fascination with trendiness of digital platforms and their users may create an over-accelerated public debate in which a disproportionate share of media attention is captured by matters which are incapable to sustain it. As the shoaling of sea waves is associated with the entering in shallower waters, so junk news bubbles may be associated with a shallower public debate, a risk that raises concerns and normative implications different from those associated with filter bubbles and fake news. While the latter can be (and has been) addressed by tweaking the recommendation algorithms to favour mainstream sources of information, this solution does not necessarily solve the problem highlighted in this paper.

Describing an attention regime that is increasingly pervasive in online media, junk news bubbles cannot be fought by censoring specific content or specific sources, but demands a deep restructuring of the system of incentives that characterize digital communication. Until social media will obtain the largest share of their profits from selling metrics of shallow and ephemeral engagement (e.g. impressions, views, clicks and shares) and until their users will be rewarded according to the same metrics, little are the chances to avoid dynamics of hyper-acceleration. This does not mean that all content producers will play the game of click bait and junk news – think of the many amazing works produced on YouTube by both mainstream and native creators (Snickars & Vonderau, 2009; Burgess & Green, 2009) – nor that online platforms can only promote superficial forms of engagement – think of how Twitter has been invested by all sorts of political activists (Gerbaudo, 2012). It does mean, however, that in the old opposition between a distracted public opinion (Lippmann, 1922, 1927) and engaged public inquiry (Dewey, 1927), accelerated attention regimes stack the odds against the latter.